\documentclass[iop,12pt]{emulateapj}
\usepackage{natbib} 
\usepackage{amssymb,amsmath}
\usepackage{amsfonts}
\usepackage{aas_macros}
\usepackage[dvipsnames]{color}
\def\km{\,{\rm km}}
\def\refnew#1{\,(\ref{#1})}

\def\D{\Delta}

\def\kcon{{k_{\mathrm{con}}}}
\def\krad{{k_{\mathrm{rad}}}}
\def\keff{k_{\mathrm{eff}}}

\def\keffr{k_{\mathrm{rad}}}

\def\kappaeffc{\kappa_{\mathrm{con}}}
\def\kappaeffr{\kappa_{\mathrm{rad}}}
\def\Gammac{\Gamma_{\mathrm{con}}}
\def\Gammar{\Gamma_{\mathrm{rad}}}
\def\drhat{\delta\hat r}

\def\epsy{\epsilon_Y}
\def\gpcc{\,\mathrm{g\,cm^{-3}}}
\def\dynpcms{\,\mathrm{dyne\, cm^{-2}}}
\def\ergpcmsK{\,{\rm erg\, cm^{-1}\, s^{-1}\, K^{-1}\,}}

\def\cm{\,\mathrm{cm}}
\def\m{\,\mathrm{m}}
\def\micron{\, \mu\mathrm{m}}
\def\rhat{\hat r}
\def\drhat{\delta\hat r}
\def\h{\,\mathrm{h}}
\def\ergpccpK{\,\mathrm{erg\, cm^{-3}\, K^{-1}}}

\def\gammaunit{\,\mathrm{erg\, s^{-1/2}\, cm^{-2}\, K^{-1}}}
\def\Gammaunit{\,\mathrm{J\, s^{-1/2}\, m^{-2}\, K^{-1}}}

\def\K{\,\mathrm{K}}

\def\su{\mathrm s}

\def\rhobar{\overline{\rho}}

\begin{document}

\title{Thermal Conductivity Of Rubble Piles}
\author{Jing Luan \& Peter Goldreich}
\email{jingluan@caltech.edu}
\affiliation{California Institute of Technology}

{\large \begin{abstract}
Rubble piles are a common feature of solar system bodies.  They are composed of monolithic elements of ice or rock bound by gravity.  Voids occupy a significant fraction of the volume of a rubble pile. They can exist up to pressure $P\approx \epsy\mu$, where $\epsy$ is the monolithic material's yield strain and $\mu$ its rigidity.  At low $P$, contacts between neighboring elements are confined to a small fraction of their surface areas.  As a result, the effective thermal conductivity of a rubble pile, $\kcon\approx k(P/(\epsy\mu))^{1/2}$, can be orders of magnitude smaller than, $k$, the thermal conductivity of its monolithic elements.  In a fluid-free environment, only radiation can transfer energy across voids. It contributes an additional component, $\krad=16\ell\sigma T^3/3$, to the total effective conductivity, $\keff=\kcon +\krad$. Here $\ell$, the inverse of the opacity per unit volume, is of order the size of the elements and voids.  An important distinction between $\kcon$ and $\krad$ is that the former is independent of the size of the elements whereas the latter is proportional to it.  Our expression for $\keff$ 
provides a good fit to the depth dependence of thermal conductivity in the top $140\,\mathrm{cm}$ of the lunar regolith.  It also offers a good starting point for detailed modeling of thermal inertias for asteroids and satellites.   Measurement of the response of surface temperature to variable insolation is a valuable diagnostic of a regolith.  There is an opportunity for careful experiments under controlled laboratory conditions to test models of thermal conductivity such as the one we outline.

\end{abstract}

\maketitle

\section{Introduction}
Observational evidence for rubble piles is varied. Mean densities, $\rhobar$, below $1\gpcc$ are typical of Saturn's icy satellites with radii smaller than $140\km$ \citep{Dougherty}, indicating porosity throughout their entire bodies.  Lunar seismometers detected strong scattering in the upper $20\km$ of the Moon implying the presence of fractures and voids \citep{Heiken}. The sharp decline in the number of asteroids with spin periods below $\sim 3\h$ demonstrates both their low mean densities, $\rhobar\leq 2\gpcc$, and weak cohesion \citep{Waszczak}. Thermal responses of asteroids and satellites to time variations of the incident solar flux yield thermal inertias approximately two orders of magnitude smaller than those of monolithic materials \citep{Delbo, Howett}.  The lunar surface is covered by a layer of regolith whose density in the top $3\m$ ranges between $1$ to $2\gpcc$\citep{Heiken} and thus below that of rock, $\rho\approx 3\gpcc$.  

\cite{Goldreich} studied the elastic behavior of rubble piles. They pointed out that voids rather than cracks are the essential difference between rubble piles and monoliths. We focus on the thermal properties of rubble piles, which have broad applications.  In the fluid free environment pertinent to asteroids and satellites lacking atmospheres, voids impede heat transfer.   

Our paper is organized as follows. In Section \ref{sec:geometry}, we evaluate the condition for a rubble pile to survive and show that it is consistent with indications from observations. We derive an order-of-magnitude expression for the effective conductivity, and compare it to measurements of the thermal conductivity of the lunar regolith in Section \ref{sec:conductivity}.  Section \ref{sec:thermal-inertia} presents a model for the thermal inertia of a rubble pile and tests it against observationally determined values for the Moon and the icy Saturnian satellites.  We conclude with a short discussion in Section \ref{sec:summary}.

\section{Conditions For Existence Of Rubble Piles}\label{sec:geometry}

Consider a typical monolithic element composing a rubble pile.\footnote{This section applies and extends aspects of the \cite{Hertz} theory of contact between elastic bodies.}  Its surface is generally coarse, i.e., covered by nubs spanning a wide range of radii of curvature \citep{Johnson}.  Contacts between neighboring elements typically involve nubs with the smallest radii of curvature, $\hat r$, that can survive strains generated by the weight of the overlying material. Smaller nubs are crushed and flattened. Figure \refnew{fig:geometry} illustrates the geometry.  
\begin{figure}
\includegraphics[width=0.9\linewidth]{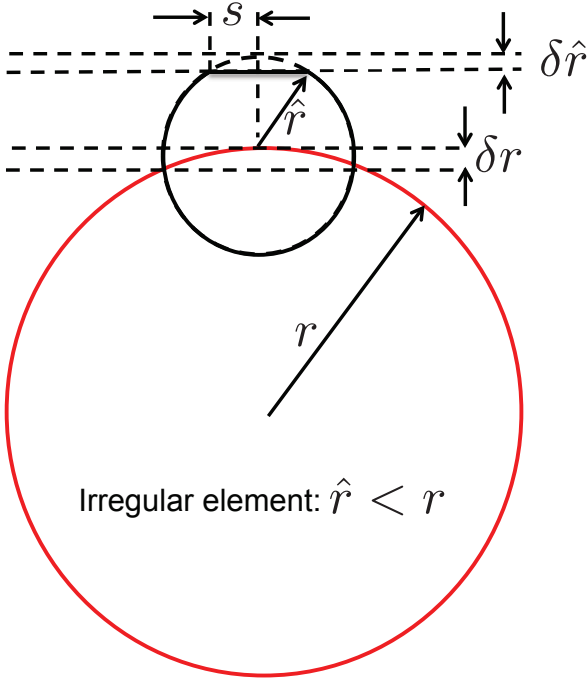}
\caption[Element geometry]{\label{fig:geometry} The geometry of a rough-surface element.}
\end{figure}

Although each element participates in several contacts, it suffices to focus on just one.  As shown in Figure \refnew{fig:geometry}, the nub is compressed by $\drhat$ resulting in contact area $s^2\sim \rhat\drhat$. Near contact, the maximum strain, $\epsy\sim \drhat/s$, and stress, $\sigma\sim \mu\epsy$, concentrate in a volume $\sim s^3$ where $\epsy$ and $\mu$ denote yield strain and shear modulus, respectively.  

At average pressure $P$, each element must transmit force $\sim r^2P$ across contact area $\sim s^2$. 
Combining the above relations, we obtain
\begin{eqnarray}
\hat r&\sim & r \left(P\over\epsy^3\mu\right)^{1/2}\, ,\label{eq:rhat}\\
s &\sim & r \left(P \over\epsy\mu\right)^{1/2}\, ,\label{eq:s}\\
\drhat &\sim & r\left(\epsy P\over \mu\right)^{1/2}\, ,\label{eq:drhat}
\end{eqnarray}
which apply for both $\rhat<r$ and $\rhat>r$. At $P\sim\epsy^3\mu$, $\rhat\sim r$. Above this pressure voids shrink leading the mean density to grow.  At $P\sim\epsy\mu$, $s\sim r$, i.e., the strain reaches $\epsy$ throughout the element, voids close and the density approaches its monolithic value.  We refer to $P<\epsy^3\mu$ as the low-pressure regime and $\epsy^3\mu<P<\epsy\mu$ as the high-pressure regime. Rubble piles do not exist at $P>\epsy\mu$. 

The hydrostatic pressure at the center of a homogeneous sphere is $P_c=(2\pi/3)G\rhobar^2 R^2$. Thus the low-pressure regime
would apply throughout bodies with $R\leq R_*$;
\begin{equation}
R_*\sim \left(\mu\epsy^3\over G\rhobar^2\right)^{1/2}\sim 10\left(\epsy\over 0.01\right)^{3/2}\km\, ,
\label{eq:Rstar}
\end{equation}
whereas the limit of the high-pressure regime would be reached at the centers of bodies with $R_{\rm max}$;
\begin{equation}
R_{\rm max}\sim \left(\mu\epsy\over G\rhobar^2\right)^{1/2}\sim 10^3\left(\epsy\over 0.01\right)\km\, .
\label{eq:Rmax}
\end{equation}
Values of $R_*$ and $R_{\rm max}$ apply to bodies composed of either rock or ice; $\mu$ is about 10 times larger for rock than for ice, but $\rho$ for ice is about 3 times smaller than that for rock.

Mean densities as small as half that of the monolithic density of their constituents would be restricted to $R<R_*$ although less substantial under-densities could persist up to $R=R_{\rm max}$.  
Rubble piles may exist in the upper layers of bodies larger than $R_{\max}$.  Given surface gravity $g$, low-pressure and high-pressure regimes would extend to depths $d_*\approx \epsy^3\mu/(\rhobar g)$ and $d_{\rm max}\approx \epsy\mu/(\rhobar g)$.  Assuming
$\epsy=0.01$ and parameters appropriate to the Moon, $g\approx 160\,\mathrm{cm\, s^{-2}}$, $\rhobar\approx 3\gpcc$, and $\mu\approx 5\times 10^{11}\dynpcms$,  we estimate $d_*\approx 10\m$ and $d_{\rm max}\approx 100\km$.  Passive seismic experiments on the moon indicate that wave scattering is strongest in the upper $20\km$ \citep{Heiken}, presumably where fractures and voids are most abundant.

\section{Effective Conductivity}\label{sec:conductivity}

\subsection{Phonon Conductivity}
Suppose the temperature drops by $\sim \Delta T$ across an element. In steady state, with uniform monolithic conductivity, $k$, and without heat sources or sinks, the temperature, $T$, satisfies $\nabla^2 T=0$.  Like the stress, the magnitude of the temperature gradient peaks in the vicinity of the contact.  Away from the contact, the conductive flux and hence ${\mathbf \nabla} T$ diminish roughly quadratically with distance implying $|{\mathbf \nabla}T|\approx \D T /s$ within distance $s$ from the contact \citep{Batchelor}.  Thus the total conductive luminosity passing through the contact is $\sim k\Delta T s$, from which we deduce that the effective conductivity
\begin{equation}\label{eq:kcon}
\kcon\sim k {s\over r} \approx k\left(P\over \epsy\mu\right)^{1/2}\, .
\end{equation}
Equation \refnew{eq:kcon} applies in both low and high pressure regimes. 

In the low pressure limit, $P<\epsy^3\mu$, perfect elastic spheres would have
\begin{equation}
\kcon\approx k\left(P\over \mu\right)^{1/3}>k\left(P\over \epsy\mu\right)^{1/2}\, ,
\label{eq:keffcsph}
\end{equation}
but measurements of $\kcon$ in granular media consisting of commercially manufactured spheres \citep{Watson64, Huetter2008} obtain results similar to those found using crushed materials of similar composition and size.  
Presumably, even the surfaces of carefully manufactured spheres possess a spectrum of small scale irregularities. 

\subsection{Photon Conductivity} \label{subsec:photoncond}
Radiation contributes to the effective conductivity in three ways.  
\begin{itemize}
\item In a fluid free environment and absent physical contacts, only radiation can transfer energy between elements.  To assess the rate at which it does so, we consider the simple setup displayed in Figure \ref{fig:slab} in which parallel monolithic slabs of thickness $\ell$ are separated by a vacuum gap of thickness $d$.  We assume steady state conditions and slabs opaque to thermal radiation.  For $0<\delta T<\Delta T<< T$, the flux, $F$ satisfies
\begin{equation}
F=4\sigma T^3\delta T \, ,
\end{equation}
\begin{equation}
F=\frac{k(\Delta T-\delta T)}{\ell} \, ,
\end{equation}
\begin{equation}
F=\krad\frac{\Delta T}{\ell+h}.
\end{equation}
These equations yield
\begin{equation}
\frac{\krad}{k}=\frac{4(\ell+h)\sigma T^3}{k+4\ell\sigma T^3}\, 
\end{equation}
\begin{equation}
\frac{\delta T}{\Delta T}=\frac{k}{k+4\ell\sigma T^3}\, 
\end{equation}
In the limit most important to our investigation, $k\gg 4\ell\sigma T^3$, $\krad\approx 4(\ell+h)\sigma T^3$ and
$\delta T\approx \Delta T$. In other words, each slab is nearly isothermal and the conductive flux passing through it is determined by the rate at which radiation transfers energy across the vacuum gap separating adjacent slabs.  Moreover, for $h\to 0$, $\krad\to 4\ell\sigma T^3$.
\item Next imagine cutting holes of radius $\sim \ell$ in random locations through each slab. Now some photons may travel a vertical distance $\sim (\ell+h)$ before striking a slab. This provides a direct radiative conductivity $\sim (\ell+h)\sigma T^3$.
\item Radiation can also transport energy through elements composed of transparent materials but we neglect this effect because ice and rock are opaque to infrared radiation.
\end{itemize}
\begin{figure}
\includegraphics[width=0.9\linewidth]{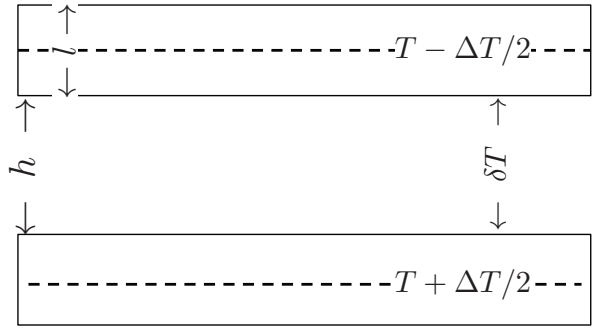}
\caption{\label{fig:slab} The central temperatures of the neighboring slabs differ by $\D T$.  Each slab has thickness $\ell$ and is separated from its neighbor by a vacuum of thickness $d$. the lower surface of the upper slab is a distance $d$ above the upper surface of the lower slab, and the temperatures of these surfaces differ by $\delta T$. }
\end{figure}

Combining the results of the first two items discussed above, we define the radiative conductivity
\begin{equation}
\krad=\frac{16\ell\sigma T^3}{3}\, ,
\label{eq:krad}
\end{equation}
where following standard convention, $\ell$ is the inverse opacity per unit volume.  In practice we expect that
$\ell\sim r$, the typical linear size of both elements and voids.

\subsection{Application to the Lunar Regolith}
\cite{Keihm}\footnote{ These results may suffer from poor thermal linkage in the probe-borestem system as remarked by \cite{Langseth}, in which $k$ was estimated to be $30\%\sim 50\%$ smaller than that in \cite{Keihm}. } determined the thermal conductivity of the lunar regolith from temperatures measured by thermocouples the Apollo 15 astronauts placed on the Moon's surface. Their results are plotted as solid circles in Figure \ref{fig:Apollo-kcond}.  The effective conductivity, $k_{\rm eff} =\kcon+\keffr$, includes contributions from both phonon and photon diffusion.  Because the latter is proportional to $T^3$, it is important to note that the values of $k_{\rm eff}$ are those appropriate for the mean temperature at each depth.  Although temporal variations in surface temperature are large, spanning most of the range between $100-400\K$, those of the mean temperature are modest, rising from $\approx 200\K$ at the surface to $\approx 250\K$
at the maximum depth of $140\cm$.  
\begin{figure}
\includegraphics[width=0.9\linewidth]{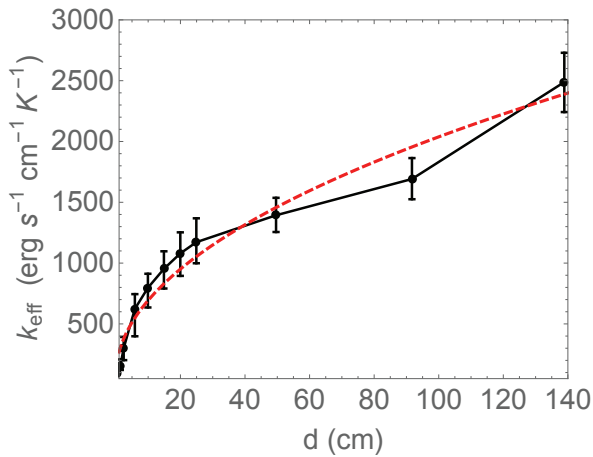}
\caption[Thermal conductivity of lunar regolith]{\label{fig:Apollo-kcond} Filled circles with error bars are values of $\keff$ as a function of depth, $d$, at the Apollo landing site \citep{Keihm}.  Our 2 parameter fit to the data given by Equation \refnew{eq:Apollocond} is shown by the dashed red line. See main text for details.} 
\end{figure}
\bigskip
Thus we neglect the latter and make a 2 parameter fit to the data inspired by contributions to $\keff$ from phonon and photon diffusion expressed by equations (\ref{eq:keffcsph}) and (\ref{eq:krad}), respectively.  We find
\begin{equation}
k_{\rm eff}=\left[197\left(d\over\cm\right)^{1/2}+72\right]\ergpcmsK\, ,
\label{eq:Apollocond}
\end{equation}
where $d$ denotes depth below the surface.  In choosing such a simple form, we are ignoring not only the depth dependence of the mean temperature but also potential variations with depth of grain size and composition.  Equating the form of $\kcon$ from equation \refnew{eq:kcon} to $197(d/\cm)^{1/2}\ergpcmsK$, yields $\epsy\approx 5\times 10^{-4}$ for $P\simeq \rhobar g d$, with $\rhobar\approx 1.5\gpcc$, $k\sim 2\times 10^5\ergpcmsK$, $\mu\approx 5\times 10^{11}\dynpcms$, and $g\approx 160\,\mathrm{cm/s^{2}}$.\footnote{Given the order of magnitude nature of our analysis, the value of $\epsy\approx 5\times 10^{-4}$ should be viewed as unremarkable. To fit the pressure dependence of the shear velocity in sand, \cite{Goldreich} require $\epsy\simeq 0.17$.}  Setting $\krad=72\ergpcmsK$ and appealing to equation \refnew{eq:krad}, we obtain $\ell\approx 10^2\,(T/220\K)^3\micron$ for the photon diffusion length. This length is several times smaller than the depth at which phonon and photon diffusion contribute equally to $\keff$.  Another relevant comparison is with grain sizes at the Apollo 15 landing site.  As reported by \cite{Papike}, grains smaller and larger than $\approx 100\micron$ contribute similar amounts to the overall density of the regolith.

\section{Thermal Inertia}\label{sec:thermal-inertia}
A body's thermal inertia is an important diagnostic of its surface. It is deduced from the amount by which variations of surface temperature, $T_\su$, lag those of the incident radiative flux.  For monolithic material, thermal inertia is defined by
\begin{equation}
\Gamma_{\mathrm{mon}}\equiv (k\rho c_p)^{1/2}=\rho c_p \kappa^{1/2}\, ,
\label{eq:Gam}
\end{equation}
where thermal diffusivity, 
\begin{equation}
\kappa=\frac{k}{\rho c_p}\, ,
\label{eq:kappa}
\end{equation}
with $c_p$ denoting specific heat capacity.   

Provided all vibrational degrees of freedom are classically excited, common monolithic insulators share the almost universal values of $\rho c_p\approx 2\times 10^7\ergpccpK$ and $\kappa\approx 10^{-2}\,\mathrm{cm^2\, s^{-1}}$ \citep{White}.  Each nucleus contributes $3k_B$ to the heat capacity and is surrounded by an electron cloud whose volume is insensitive to the screened nuclear charge. Taking each nucleus to occupy a cube with $2.75\AA$ sides yields $\rho c_p\approx 2\times 10^7\ergpccpK$. Most of the thermal energy is stored in the shortest wavelength lattice vibrations and typically these propagate a few lattice spacings at speeds of a few $\mathrm{km\, s^{-1}}$ before being scattered. With mean free path $\lambda\approx 10\AA$ and propagation speed $v\sim 3\,\mathrm{km\, s^{-1}}$,  $\kappa\approx\lambda v/3\sim 0.01\,\mathrm{cm^2\, s^{-1}}$.  Consequently,\footnote{We express thermal inertia in the cgs unit, $\gammaunit$, instead of the more convential mks unit $\Gammaunit=10^3\gammaunit$.} 
\begin{equation}
\Gamma_{\mathrm{mon}}\approx 2\times 10^6\gammaunit \, .
\label{eq:Gammamon}
\end{equation}

We motivate the definition of thermal inertia by means of a simple example.  Consider the response of a body to the sudden imposition of a constant incident flux $F$ at $t=0$. In the interest of simplicity, we treat a monolithic body with $k$ independent of both $P$ and $T$ and assume zero albedo.  Surface temperature, $T_\su$, is determined by balancing the incident flux against the sum of the outward radiative flux and the inward conductive flux;
\begin{equation}
F=\sigma T_\su^4+k {dT\over dz}\sim \sigma T_\su^4+\left(\frac{k\rho c_p}{t}\right)^{1/2}T_\su\, \, ,
\end{equation}
where we approximate $dT/dz$ by $T_\su$ divided by the penetration depth, $(\kappa t)^{1/2}$, for thermal diffusion during time $t$. Initially most of the incident flux is conducted inward and 
\begin{equation}
T_\su\sim \left(t\over \rho kc_p\right)^{1/2}F\, .
\end{equation}
At 
\begin{equation}
t_{\rm lag}\sim \frac{k\rho c_p}{(\sigma T_\su^3)^2}=\left(\frac{\Gamma}{\sigma T_\su^3}\right)^2\, , 
\label{eq:tlag}
\end{equation}
the outward radiative and inward conductive fluxes are comparable.  Thereafter, the radiative flux dominates and $T_\su$ asymptotically approaches $(F/\sigma)^{1/4}$. 

Fitting variations of surface temperature in response to variations of incident solar flux requires a model for $\keff$.  Ours includes depth dependence from $\kcon$ and temperature dependence from $\krad$. Here we consider limiting cases in which either the former or the latter dominates.  

\begin{itemize}
\item 
Suppose phonon diffusion dominates heat transfer; $\keff\approx \kcon$. 

We evaluate $\Gammac\approx (\kcon\rhobar c_p)^{1/2}$ at the depth to which the thermal wave propagates in time $t$, $d\approx (\kcon t/(\rhobar c_p))^{1/2}$.  This procedure yields
\begin{equation}
\Gammac\approx \left(k^2\rhobar^2c_p\,g\,t^{1/2}\over\epsy\mu\right)^{1/3}\, ,
\label{eq:Gammac}
\end{equation}
for flux variations on timescale $t$.  

\item Next consider the opposite limit in which photon diffusion dominates; $\keff\approx \krad$. 

Here 
\begin{equation}\label{eq:Gammar}
\Gammar=\left(\frac{16\ell\sigma T^3\rhobar c_p}{3}\right)^{1/2}\, ,
\end{equation}
which has no explicit dependence on $t$.

\end{itemize}

From observations during a lunar eclipse with umbral duration $t\sim 2\h$, \cite{Muncey} estimates $\Gamma\approx 2.8\times10^4\gammaunit$.   \cite{Linsky} estimates $3.9<\Gamma<6.7\times 10^4\gammaunit$ from data obtained during a lunation,
$t\approx 28\,$d.  From \cite{Keihm} and our fit to $\keff$ in equation \refnew{eq:Apollocond}, it appears that both phonon and photon
diffusion contribute significantly to $\Gamma$ during a lunar eclipse but that the phonon contribution dominates during a lunation.  Unfortunately, neither these old data nor our theory are precise enough to justify a more detailed analysis. Mid-infrared measurements made by the radiometer on the Lunar Reconnaissance Orbiter presents a comprehensive picture of the Moon's surface temperature over 4 lunations \citep{Vasavada2012}.  It clearly warrants more careful modeling than we are currently capable of doing. 

 Microwave measurements from the Rosetta orbiter made prior to Philae's landing on comet 67P/Churyumov-Gerasimenko, 
were interpreted to imply a representative thermal inertia, $\Gamma$, in the range $(1-5)\times 10^4\gammaunit$ for the overall surface \citep{Gulkis}.  Shortly thereafter, diurnal temperature variations measured at the Philae landing site, Abydos, yielded an estimate of $\Gamma=(8.5\pm 3.5)\times 10^4\gammaunit$ for the local thermal inertia \citep{Spohn} .  These low values imply a porous surface.  Lack of knowledge of the local regolith prevents the direct application of our formulae for $\Gamma$. However, in-situ measurements of the variation of the surface temperature over the comet's orbital period might separate contributions from phonon and photon conduction because we expect $\Gammac\propto t^{1/6}$ and $\Gammar$ independent of $t$.

Thermal inertia is an essential component of the Yarkovsky effect which drives significant orbital migration of small asteroids \citep{Bottke06}. This migration is responsible for the rate at which the semi-major axes of members of asteroid families separate.  It also impacts the timescale for the delivery of meteorites from the asteroid belt to Earth.  Phase lags expressed in radians of rotational phase, $\Delta\phi$, can be estimated by multiplying $t_{\rm lag}$ in equation \refnew{eq:tlag} by the spin frequency of the asteroid;
\begin{eqnarray}\label{eq:delphi}
\Delta\phi &\approx & \frac{2\pi}{P_{\mathrm sp}}\left(\Gamma\over\sigma T^3\right)^2\sim 5\times 10^4\left(P_{\mathrm sp}\over \h\right)^{-1}\cr
&\times&\left(T\over 200\K\right)^{-6}\left(\Gamma\over\Gamma_{\mathrm{mon}}\right)^2\, .
\label{eq:Delphi}
\end{eqnarray}
Phase lags given by equation \refnew{eq:Delphi} are relevant to the diurnal Yarkovsky effect. There is also a seasonal Yarkovsky effect which involves the phase lag expressed in terms of the orbital phase.  In each case, migration rates are optimized for phase lags of order a radian \citep{Bottke06}.

\cite{Howett} estimate thermal inertias for the Saturnian satellites Mimas, Enceladus, Tethys, Dione, Rhea and Iapetus from infrared data obtained by the Cassini orbiter.  Each satellite has an outer ice shell 
and spins synchronously so incident solar flux variations occur on timescales of its orbital period, $P_{\mathrm orb}$.  Thermal inertias are plotted in Figure \refnew{fig:thermal-inertia}. They are low and show no obvious trend with orbit period.  This suggests that heat transport by photon diffusion probably dominates that by phonon diffusion at the top of their regoliths; i.e., $\kappaeffr>\kappaeffc$ even for Iapetus.  Otherwise, it would be expected from equation \refnew{eq:Gammac} that the $10^3$ fold increase of $g^2 P_{\mathrm orb}$ in going from Mimas to Iapetus would lead to a noticeable, factor of $\approx 3$, rise of $\Gamma$ as indicated by the dashed line in Figure \refnew{fig:thermal-inertia}.\footnote{Here we assume that these satellites have similar regoliths.} 

The average $\Gamma$ for Saturn's satellites is $1.3\times 10^4\gammaunit$ as shown by the dotted blue line in Figure \refnew{fig:thermal-inertia}. With $\Gammar\approx 7.8\times 10^4\,(T/100\K)^{3/2}\,(\ell/\cm)^{1/2}\gammaunit$, it follows that $\ell\approx 0.04(90\K/T)^{3}\cm$, an unremarkable value.  But an upper limit of $\Gammac\leq 1.3\times 10^4\gammaunit$ for Iapetus is
problematic. Substitution of $\rho \sim 0.93\gpcc$, $\mu\sim 4\times 10^{10}\dynpcms$, $c_p\sim 2\times 10^7\ergpccpK$,
$k\sim 2\times 10^5\,\mathrm{erg\, cm^{-1}\, s^{-1}\, K^{-1}}$, $g\approx 22.4\,\mathrm{cm\, s^{-2}}$, $\epsy=0.01$ and $t=79/(2\pi)\,\mathrm{d}$ into equation \refnew{eq:Gammac} yields $\Gammac\approx 3.5\times 10^4\gammaunit$. The astute reader will recognize that the substituted values are appropriate for water ice at $273\K$.  Moreover, although $\epsy=0.01$ is a reasonable value for the yield strength of a single crystal of cold ice, it is a large one for polycrystalline ice \citep{Hobbs, Schulson}.  Substituting parameters suitable for pure water ice at $90\K$ would not help; $k$ would be larger and although $c_p$ would be smaller, the product $k^2 c_p$ would be slightly larger. We have checked this statement for $T$ down to $173\K$ and see no reason why it would not apply for $T$ as low as $90\K$.  A significant reduction of $\rho$ is more plausible; on Earth, accumulations of dry snow with water content below 30$\%$ are common. The weaker gravity on Iapetus would permit even lower densities than on Earth.  Even at temperatures below 90$\K$, ice grains might bond to their neighbors.\footnote{Bonding increases contact areas at fixed pressure and implies finite contact areas at zero pressure \citep{Johnson}.} Bonding could enable both low $\rho$ and large $\epsy$. According to \cite{BGT84}, bonding is the most likely explanation for the paucity of sub-centimeter water ice particles in Saturn's rings.

\begin{figure}
\includegraphics[width=0.9\linewidth]{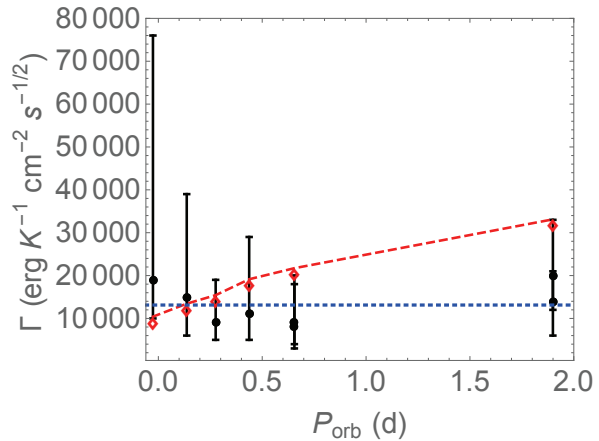}
\caption[Thermal inertia of Saturnian moons]{\label{fig:thermal-inertia} Comparison between data \citep{Howett} in black and theoretical estimates based on the assumed dominance of either phonon diffusion (dashed red line) or photon diffusion (dotted blue line). The former is normalized to pass through the lowest value for $\Gamma$ of Mimas. Its rise with increasing $P_{\mathrm orb}$ is attributable to the $\approx 10^3$-fold increase of $g^2P_{\mathrm{orb}}$ from Mimas to Iapetus. The constant value of $\Gammar\approx 1.3\times 10^4 \gammaunit$ corresponds to setting $\ell\approx 0.04\cm$ for $T\approx 90\K$.  See main text for more details. }
\end{figure}

\section{Summary}\label{sec:summary}
 Together, phonon and photon diffusion determine the thermal conductivity of a rubble pile. 

Phonons transmit heat through contacts between neighboring elements.  Conductivity due to phonon diffusion in granular materials is independent of the sizes of the individual elements. We consider irregularly shaped elements whose mutual contact areas in the low pressure regime, $P<\epsy^3\mu$, are smaller than those for spheres.  Consequently,  they provide a lower phonon conductivity than spheres, but one that increases more rapidly with pressure, $\propto P^{1/2}$, rather than, $\propto P^{1/3}$ for spheres.  At $P\approx\epsy^3\mu$, phonon conductivities of irregular elements and spheres are equal.  At still higher pressure, the contact areas of both spheres and irregularly shaped elements are governed by equation (\ref{eq:s}).  

The effective conductivity in the top $140\cm$ of the lunar
regolith as deduced from radiometer measurements of surface temperatures \citep{Keihm} roughly agrees with our prediction $\keff\propto d^{1/2}$.  But measurements of the annual temperature variation by probes analyzed by \cite{Langseth} indicate that $\keff$ is nearly constant at about $10^{-2}$ of the monolithic conductivity for $d\leq 250\cm$.  This result has no simple explanation since contact areas should monotonically increase with increasing pressure.  

Photons transmit heat across voids between elements. Photon diffusion contributes $\krad\propto \ell T^3$, where $\ell$ is the linear size of a typical element.  A subtle argument in \ref{subsec:photoncond} shows that the element size is more important than
the void size in determining the appropriate value of $\ell$.

Thermal inertias of Saturnian satellites exhibit little dependence on $g^2 P_{orb}$ suggesting that photon diffusion may dominate $\keff$.  This is surprising both because their surface temperatures are low and the implied values of $\keff$ are smaller than one might expect from phonon diffusion alone.

\section*{Acknowledgement}
We thank David Rubincam, Thomas A. Prince, and Christian D. Ott for their comments and suggestions.

\vfill\eject

\end{document}